# Integrated angstrom-tunable polarization-resolved solid-state photon sources

Yinhui Kan[1*], Paul C. V. Thrane[1,2], Xujing Liu[1], Shailesh Kumar[1], Chao Meng[1], Radu Malureanu[3,4], Sergey I. Bozhevolnyi[1*]

[1] Centre for Nano Optics, University of Southern Denmark, DK-5230 Odense M, Denmark.

[2] SINTEF Microsystems and nanotechnology, Gaustadalleen23c, 0737 Oslo, Norway.

[3] Department of Electrical and Photonics Engineering, Technical University of Denmark, DK-2800 Kongens Lyngby, Denmark.

[4] National Centre for Nano Fabrication and Characterization, Technical University of Denmark, DK-2800, Kogens Lyngby, Denmark.

*Corresponding author. Email: yk@mci.sdu.dk; seib@mci.sdu.dk

**Abstract**

The development of high-quality solid-state photon sources is essential to nano optics, quantum photonics, and related fields. A key objective of this research area is to develop tunable photon sources that not only enhance the performance but also offer dynamic functionalities. However, the realization of compact and robust photon sources with precise and wide range tunability remains a long-standing challenge. Moreover, the lack of an effective approach to integrate nanoscale photon sources with dynamic systems has hindered tunability beyond mere spectral adjustments, such as simultaneous polarization control. Here we propose a platform based on quantum emitter (QE) embedded metasurfaces (QEMS) integrated with a microelectromechanical system (MEMS)-positioned microcavity, enabling on-chip multi-degree control of solid-state photon sources. Taking advantages of MEMS-QEMS, we show that typically broadband room-temperature emission from nanodiamonds containing nitrogen-vacancy centres can be narrowed to 3.7 nm and dynamically tuned with angstrom resolution. Furthermore, we design a wavelength-polarization-multiplexed QEMS and demonstrate polarization-resolved control of the MEMS-QEMS emission in a wide wavelength range (650 nm to 700 nm) along with polarization switching at sub-millisecond timescales. We believe that the proposed MEMS-QEMS platform can be adapted for most existing QEs, significantly expanding their room-temperature capabilities and thereby enhancing their potential for advanced photonic applications.

**Main text**

Solid-state photon sources serve as essential building blocks in both fundamental quantum nanophotonics and advanced quantum technologies[1-3]. However, most solid-state quantum emitters (QEs), such as quantum dots[4,5] and colour centres in nanodiamonds[6], suffer from intrinsic limitations at room temperature, including broad bandwidth, omnidirectionality, and poorly defined polarization[1,7,8]. Over the past decade, many investigations have been concerned with engineering of existing QEs to mitigate their deleterious characteristics[9,10]. Among them, integrating QEs with micro/nano structures has emerged as one of the most attractive approaches for at-source manipulating the properties of photon emission[11-13], for example, coupling with individual nanoantennas to enhance emission rates[14–16]. Recently, on-chip integration of QE with planar subwavelength arrays (i.e., QE-coupled metasurfaces) has been proposed as 'meta-QE', offering attractive solutions for realizing versatile photon sources of circular[17,18], radial[19], and linear polarizations[20,21] with collimated and directional photon emission. Even though the meta-QE configurations demonstrated greatly improved photon emission characteristics without use of traditional bulky components, these photon sources remain static, i.e., with fixed functionalities once constructed[22].

Developing dynamically tunable photon sources is highly desirable for diverse applications, such as multidimensional, high-capacity information processing[23,24]. Approaches like electrically tuning photon sources via the Stark effect[25,26] are conceptually straightforward, but offer limited spectral tuning ranges (typically < 1nm) due to the tightly confined electronic states inherent to QEs[25]. Alternatively, QE photon emission can be modulated using movable cavities, assisted by external three-dimensional positioning systems, e.g., fiber-based parabolic-flat microcavities, to selectively couple the photon emission into a cavity mode[27] and thus modulate the Purcell enhancement[28]. However, the significant bottlenecks of this design are that the parabolic facet, with fixed curvature, can only focus one QE at a time and is hard to address other critical emission properties, e.g., the polarization remains poorly defined and lacks the tunability, hindering further practical applications[29-32]. Achieving wide-range spectral tunability and well-resolved polarization requires efficient spectral channelling of photon emission into a cavity mode, together with delicate control of multiplexed polarization states. Nevertheless, this remains a challenge due to the spatial-temporal constraints of manipulating the coupling of omnidirectional and wide-cone photon emission from an unmodified nanoscale point source within a cavity. Overall, chip-scale integrated tunable photon sources with precise, wide-range, and fast tunability remain elusive and not yet accessible by state-of-the-art micro/nano-scaled photon sources.

In this work, we propose an on-chip platform integrating QE-embedded metasurfaces (QEMS) with a microelectromechanical system (MEMS)-tuned microcavity for realizing dynamic polarization-resolved photon emission, featuring real-time, precise, and wide-range wavelength tunability and polarization switching. MEMS-QEMS takes advantage of both QEMS, spatially converting QE-excited surface plasmon polaritons (SPPs) into unidirectional polarized photon emission, and MEMS-controlled distributed Bragg reflector (DBR), tuning the microcavity resonance wavelength to temporally modulate the photon emission spectrum. Following QEMS nanofabrication and MEMS-QEMS assembly, we show that typically broadband (~ 100 nm) room-temperature emission of nitrogen vacancy centres in nanodiamonds (NVs-NDs) can be narrowed to 3.7 nm (quality factor of 180). By electrically actuating the MEMS-DBR microcavity, the photon emission spectrum is dynamically and precisely modulated with an angstrom level tunability. Furthermore, we design a wavelength-polarization-multiplexed QEMS capable of encoding different polarizations at different wavelengths, e.g., $x$ linear polarization at 700 nm and $y$ linear polarization at 650 nm. We demonstrate polarization-resolved control of MEMS-QEMS emission in a wide wavelength range (650 nm to 700 nm) along with polarization switching, whose temporal response is at sub-millisecond timescales (< 0.8 ms). Our work establishes a new paradigm for advanced solid-state photon sources, facilitating a series of intriguing functionalities, such as spatiotemporal polarization-wavelength multiplexing for applications in nano optics and quantum photonics.

**Device operation principle**

The proposed MEMS-QEMS integrates QEMS into a dynamic Fabry–Pérot (FP) microcavity, constructed using a DBR and a MEMS-actuated mirror (Fig. 1a). The DBR is designed with 5 pairs of $SiO_2/TiO_2$ layers with thicknesses of 101 nm/82 nm (Supplementary Fig. 1), simultaneously allowing for high transmission around 532 nm (as an excitation window, Fig. 1b) and high reflection within the DBR region (625 nm - 725 nm), which respectively corresponds to the excitation laser and the main emission wavelengths of NVs-ND at room temperature[17]. By varying the gap distance $d$ of the microcavity, the resonance wavelength shifts accordingly, as illustrated in Fig. 1c with the calculated reflectance of the microcavity as a function of the gap distance. The gap distance, i.e., the microcavity length, can be controllably adjusted by applying actuation voltages to a piezoelectric membrane that suspends the MEMS mirror. Thus, the FP resonance wavelength can be modulated. The dashed lines represent experimental results for different voltages applied to the MEMS (Supplementary Figs. 2 and 3 and Supplementary Table 1), while the dots mark the dips in the measured spectra, which closely match the calculations and predict the corresponding gap distances. For example, at the actuation voltage Vc, one of the dips in the spectrum is around 670 nm, where the electric field is highly confined within the cavity (on-resonance, Fig. 1d), whereas the light is not confined for off-resonance wavelengths (e.g., at 650 nm). This provides a possibility for dynamically tuning photon sources by spectrally channelling QE emission into the MEMS cavity.

The MEMS mirror consists of a 20 nm thick $SiO_2$ layer covered silver film (thickness of 150 nm), deposited on one side of a MEMS chip made from a silicon-on-insulator (SOI) wafer and a thin piezoelectric film. On top of the MEMS mirror, versatile QEMS can be constructed based on the well-developed fabrication process in which the positions of NDs can be precisely determined prior to surrounding nanostructure fabrication[17]. The QEMS are ultracompact, featuring the overall size of 17 µm and the thickness of 150 nm (Fig. 1e). The NVs-ND is precisely positioned at the centre of the configuration and excited by a tightly focused radially polarized 532 nm laser that efficiently penetrates the DBR (Fig. 1a, b, and f). The use of radially polarized laser beam facilitates the excitation of NVs-ND featuring a vertical electric dipole, which can efficiently excite SPPs propagating on the surface ($x$-o-$y$ plane). Designed nanostructures around the QE can spatially convert QE-excited SPPs into unidirectional and collimated photon emission with desired polarizations (Fig. 1f). With a well-designed metasurface the divergence angle of the unidirectional photon emission can be reduced to below 3º (red line in Fig. 1g), significantly enhancing its coupling with the microcavity mode compared to the uncoupled QE (blue line in Fig. 1g, magnified 100 times to make it visible), thereby eliminating the need for parabolic mirrors in conventional designs[27,28]. After fabricating the QEMS, the structured MEMS mirror is mounted with a DBR and subsequently wire-bonded to a printed circuit board (PCB) to apply voltages (Extended Data Fig. 1). Figure 1h shows the back and front sides of the device. Two types of QEMS are fabricated: one is the QE-coupled Bullseye grating (considered as a trivial case), and the other is the polarization-multiplexed QEMS, as shown in the dark field microscope image taken in front of the DBR. We would like to note that, although MEMS have recently been integrated with optical metasurfaces[33–35], these implementations were primarily designed for modulating free-space classical light. Additionally, these MEMS have primarily functioned as moveable mirrors reflecting external incident monochromatic light from outside of the device, without enabling dynamic spectral tuning. In contrast, the MEMS-QEMS system proposed here allows one to simultaneously enhance the performance of nanoscale QE and offer dynamic functionalities.

**Narrow-bandwidth and angstrom-level tunability**

To validate the proposed MEMS-QEMS, we use Bullseye gratings to on-chip couple with NVs-NDs, forming a simple type of meta-QE, which are then integrated with the MEMS-DBR microcavity (Fig. 2a). The Bullseye configuration efficiently converts QE-excited SPPs into unidirectional photon emission normal to the surface[19], thereby facilitating the coupling of emitted photons into cavity modes compared to uncoupled QEs. The Bullseye dielectric grating is made of hydrogen silsesquioxane (HSQ) with a refractive index of 1.41 and the grating period is set as 550 nm, targeting the emission wavelength around 670 nm. As shown in the inset of Fig. 2a, the ND is positioned at the centre of the Bullseye grating, which is essential for generating photon emission with the desired direction and polarization[17,19]. By appropriately setting the voltages to adjust the gap distance of microcavity, the full width at half maximum (FWHM) of the photon emission is narrowed to 3.7 nm, with the quality factor being 180 (at 670 nm). This narrow emission offers great advantages in comparison with typical broadband (about 100 nm) emission of NVs-NDs at room temperature without coupling to the MEMS-DBR cavity (shaded area in Fig. 2b). Moreover, it is possible to further narrow the bandwidth of photon emission by increasing the number of DBR pairs, e.g., the FWHM would be down to 0.38 nm with 10 pairs of DBR, while it should balance between the intensity and quality factor of photon emission. Future optimizations could involve replacing the Ag mirror with a dielectric multilayer mirror to minimize absorption losses. This change would also allow QE-excited surface waves to transition from SPPs to Bloch waves, eliminating plasmonic losses and further enhancing the quantum efficiency. In the current SPP-based configuration, using high-index dielectric materials[36], for example $TiO_2$, to build the QEMS would decrease the influence of SPP absorption and thereby increase the brightness and overall efficiency.

The far-field emission pattern is a collimated donut-shape with a divergence angle of 7º. When characterizing with a polarizer and rotating its orientation (Extended Data Fig. 2), two lobes always appear, demonstrating the radial polarization state of the generated photon emission. For this MEMS-QEMS, the estimated gap distance can be tuned from 2.3 μm to 4.2 μm (Supplementary Table 1) when varying the voltage by 23 V, most of which compensates for non-parallel initial alignment. With more advanced packaging and alignment techniques to assemble the chip, we expect the cavity length can be reduced to hundreds of nanometres and the same tunability to be achievable with around 2V. The emission peak wavelength is about 670 nm when voltage Vc is applied, with an estimated gap distance of 2.495 μm. By varying the voltage with a step of 250 mV, we show that the emission peak gradually shifts from 669.15 nm to 670.71 nm. Furthermore, by fine-tuning the voltage in 50 mV steps, we report the emission peak can be adjusted with angstrom-level precision. The measured results reveal a quasi-linear relationship between the actuation voltage and the emission peak wavelength (Extended Data Fig. 3), featuring an average shift of 0.8 Å per 50 mV. This fine precision is achieved owing to several key properties of the proposed MEMS-QEMS system. First, the MEMS mirror features an ultra-flat surface, with a thick (400 μm) silicon mass coated by Ag ensuring exceptional planarity[37]; it also incorporates tip/tilt motion in addition to up/down, allowing precise alignment and high parallelism between the MEMS-Ag and Bragg mirrors. Second, the emission from the QEMS is highly collimated in the direction normal to the mirror surface, thereby maintaining a stable and uniform effective cavity length. Third, the MEMS system enables both long-range and highly precise control over the cavity length through piezoelectric thin-film membranes, enabling sub-nanometer displacements with mV control voltage adjustments. While this precision is remarkable, we note that the maximum precision of the MEMS-QEMS may still not have reached, primarily due to the resolution limitation of the spectrometer used in the measurements.

**Wavelength-polarization-multiplexed QEMS**

The functionality of MEMS-QEMS can be extended to develop versatile dynamic photon sources that go beyond mere spectral tunability—for example, enabling encoding with different polarizations, a capability not achieved by other technologies (Supplementary Table 2 and Table 3). However, even for the static scenario, most of state-of-the-art photon sources are designed to operate at a single wavelength with specific polarization states[17,20] or vortex modes[18,38]. Realization of compact solid-state photon sources capable of emitting photon emission with different polarizations at different wavelengths has long remained elusive. Therefore, to develop dynamically tunable polarized photon sources, we propose, as the first step, a wavelength-polarization-multiplexed QEMS that fully utilizes spatial freedom with meticulously designed local element and global arrangement. We use anisotropic dielectric (HSQ) bricks with a width of 100 nm, a length of 350 nm, and a height of 150 nm as the basic elements. By appropriately rotating the elements at specific angles, the QE-excited radially polarized SPPs can be outcoupled into high-purity linearly polarized free-space photon emission[20]. To generate photon emission with different polarizations, such as linear polarization in the $x$ or $y$ direction (LPx or LPy), the metasurfaces should be spatially arranged in corresponding regions. Unlike previous works that focus solely on polarization control, our configuration allows for simultaneous encoding of spatially resolved polarizations at different wavelengths, a complex functionality whose implementation requires meticulous design considerations taking into account the relationship between the array period and wavelength of outcoupled photon emission. As shown in Fig. 3a, by increasing the period of the array, the peak wavelength of outcoupled photon emission increases correspondingly, following a quasi-linear relationship.

Based on this relationship, two metasurfaces (M1 and M2) are designed with periods $P_1$ = 560 nm and $P_2$ = 655 nm for photon emission wavelengths of 650 nm and 700 nm, respectively. Note that the periods have been slightly detuned from the theoretical estimation (Fig. 3a) to increase the spatial separation between two polarisations. The performances of M1 and M2 for different polarizations and different wavelengths are simulated using three-dimensional finite-difference time-domain (FDTD) method. As illustrated in Fig. 3b, for M1 ($P_1$ = 560 nm), the photon emission is dominated by LPy polarization state. For the intensity of the emission spot, it decreases from the short wavelength to the long wavelength, i.e., from 1 (at 650 nm) to 0.004 (at 700 nm). A reverse trend is observed for M2 ($P_2$ = 655 nm), and conversely, the dominant polarization is LPx, as the metasurface is arranged in the $x$ direction. Furthermore, we combine these two parts into a composite metasurface and fabricate it around a preselected NVs-ND, as shown in the colorized scanning electron microscopy (SEM) image in Fig. 3c. From the far-field emission patterns in both simulation and experiment, at 650 nm, a bright collimated spot appears for LPy. Although there are two dim lobes for the LPx channel, their intensities are relatively weak. Most importantly, the photon emission of LPy and LPx are spatially separated, which is crucial for selectively extracting polarized photons at specific wavelengths. A similar phenomenon is observed at the wavelength of 700 nm, while it has a reverse polarization with the bright spot of LPx and spatially separated LPy dim lobes. This design strategy can also be applied to narrowband photon emission. For example, well-resolved spatial polarization encoding has been designed for emission bandwidths as narrow as 15 nm (660 nm and 675 nm, respectively) (Supplementary Fig. 4). Moreover, the proposed approach can be extended beyond the two-channel configuration. For instance, by segmenting the in-plane area into three distinct regions, it is possible to encode three different polarization states at different wavelengths (Supplementary Fig. 5).

**Wide range tunability and polarization switching**

With the wavelength-polarization-multiplexed QEMS at hand, we further integrate it with the MEMS-DBR microcavity to achieve wide-range tunability and dynamic polarization control (Fig. 4a). By applying an actuation voltage Vb (Fig. 1c and Supplementary Table 1), the gap distance is tuned close to 2.397 um (Supplementary Fig. 2), resulting in photon emission at 650 nm (Fig. 4b). In this regime, the LPy component is significantly larger than LPx, attributed to the efficient conversion of QE-excited SPPs into LPy photons by the M1 part of multiplexed QEMS at this wavelength. Two minor peaks, near 590 nm and 725 nm, are related to other resonance cavity modes (Supplementary Fig. 2), which in principle can be suppressed by reducing the gap distance, for instance, to 1.1 μm, and thereby enabling a single dominant emission peak within the NVs-ND emission wavelength range. When a different actuation voltage Vd (Fig. 1c and Supplementary Table 1) is applied, the estimated gap distance is adjusted to around 2.641 μm (Supplementary Fig. 2), shifting the emission peak to 700 nm that is dominated by LPx (Fig. 4c). In this regime, the outcoupled photon emission is governed by the M2 part of multiplexed QEMS, which is associated with LPx (Fig. 3b). The spectral performances coincide with the measured far-field emission patterns, where a bright spot is observed at 650 nm with LPy (Fig. 4d) under the actuation voltage Vb, while a bright spot is observed at 700 nm with LP$x$ (Fig. 4e), when the actuation voltage Vd is applied.

Additionally, the proposed MEMS-QEMS is not limited to these primary regimes. It can be tuned to other wavelengths, for example to 670 nm, where the LPx and LPy components are comparable (Extended Data Fig. 4). Moreover, beyond single-emission peak, simultaneous dual-emission peaks at both 650 nm and 700 nm can also be achieved by adjusting the gap distance to around 3.689 μm (Extended Data Fig. 5). The MEMS-QEMS exploits the multiple degree of spatial freedom: utilizing the *x*-*y* plane for encoding multiplexed polarization and wavelength via on-chip QEMS, while utilizing the *z* direction for dynamically tuning the microcavity to enable temporally modulated and spectrally controlled photon emission. It provides a new design strategy for dynamically on-chip manipulating photon emission of solid-state photon sources, different from previous parabolic microcavities[27,28] and a recent work using MEMS for on-chip control of two-dimensional materials[39]. By actuating the MEMS-DBR cavity with a periodic rectangle signal and detecting the different wavelength photons, fast switching between two regimes is observed, with the rise/fall times of ~0.75/0.45 and ~0.64/0.18 ms, respectively. The lifetimes of photon sources in both regimes are similar (19.0 ns and 17.8 ns), as the change of Purcell factor is modest when the estimated gap distance shifts from 2.397 μm to 2.641 μm (Supplementary Fig. 6). Moreover, stability measurements under relatively high excitation power (1 mW) over 5 minutes confirm the robustness of the MEMS-QEMS (Supplementary Fig. 7). The device is stable at room temperature and does not require cryogenic conditions. On the MEMS side, extensive research in recent years has significantly improved the platform reliability, with demonstrated high stability across different temperatures[40]. Furthermore, the potential degradation of Ag mirrors can be mitigated through proper passivation techniques or by replacing Ag with dielectric films to enhance longevity.

**Conclusions**

We propose a MEMS-QEMS platform that integrates QEMS with a MEMS-DBR tunable microcavity, offering unprecedented control over solid-state photon sources. We demonstrate the capability to shape the typical broadband NVs-ND photon emission into a narrow bandwidth of 3.7 nm with a quality factor of 180, using a Bullseye configuration to generate a radial polarization state. By varying actuation voltages to the MEMS-DBR microcavity, dynamically tunable photon emission from solid-state photon sources has been demonstrated with an angstrom-level precision. Furthermore, we propose a wavelength-polarization-multiplexed QEMS to encode different linear polarizations at different wavelengths. This enables wide range tunability of photon emission together with polarization control, e.g., dynamic switching of photon emission between 650 nm with LPy and 700 nm with LPx. Compared to state-of-the-art tunable photon sources (Supplementary Table 2 and Table 3), the proposed MEMS-QEMS platform demonstrates significant advancements in spectral tuning range and precision, chip-scale integration, and polarization control and tunability. Beyond polarization-wavelength multiplexing, the proposed MEMS-QEMS platform can be extended further and applied to design versatile dynamic photon sources. Note that any QEMS generating well collimated and propagating normal to the surface photon beams can be incorporated in the proposed MEMS-QEMS platform. For instance, integrating vortex-generating QEMS[38,41] would facilitate the realization of dynamically tunable photon sources generating multiple vortices. Compact photon-source configurations with diverse and easily incorporated functionalities can widely be applied in modern quantum photonic technologies, for example for generating quantum high-dimensional structured light[23,24].

In this work, we utilize the NVs-ND as a prototype, leveraging their broadband emission at room temperature. We note that MEMS-QEMS can readily be applied to a wide range of existing QEs[1]. By carefully designing the QEMS, it is possible to adapt this approach for narrowband QEs, such as semiconductor nanocrystals[42]. The concept can also be extended to simultaneously incorporate diverse solid-state QEs when building QEMS, such as tin-vacancy centres ND emitting at 620 nm[43] and silicon-vacancy centres ND emitting at 738 nm[44], enabling the realization of dynamic multi-wavelength photon sources. This is feasible from a technical perspective, for example, by incorporating advanced deterministic positioning techniques for QEs [45,46], which can simplify the fabrication process for complex QEMS and further improve the performance. The MEMS-QEMS shows relatively fast switching between two operational regimes, with response times at sub-millisecond timescales (< 0.8 ms). According to previous works where MEMS was utilized as tunable mirrors to turn the metasurfaces on or off for free-space light, the response time is primarily determined by the specific type of MEMS employed[33,34]. By using faster MEMS[35,47] together with long lifetime QEs[48], it is possible to realize spatiotemporal photon sources, which could be harnessed for advanced applications, such as high-capacity information processing[49,50] and quantum cryptography[51]. We believe that the proposed platform provides a new avenue for significantly enhancing the capabilities of solid-state QEs, offering exciting opportunities in nano optics, quantum photonics, and related fields.

**Acknowledgments:** We gratefully acknowledge Torgom Yezekyan, Osamu Takayama, and Vladimir Zenin for assistance with the experiment. The authors acknowledge the support from European Union's Horizon Europe research and innovation programme under the Marie Skłodowska-Curie Action grant 101064471 (YHK), Research Council of Norway grant 323322 (PCVT), Villum Fonden 50343 (CM), and Villum Kann Rasmussen Foundation (Award in Technical and Natural Sciences 2019) (SIB).

**Author contributions:** Y.H.K. and S.I.B. conceived the idea. Y.H.K. performed the theoretical modelling. Y.H.K., P.C.V.T., X.J.L., and R.M. fabricated the samples. Y.H.K. and X.J.L. with assistance from S.K. and C.M. performed the experimental measurements. Y.H.K., S.K., X.J.L., P.C.V.T., C.M., and S.I.B. analysed the data. S.I.B. and Y.H.K. supervised the project. Y.H.K. wrote the manuscript with contributions from all authors.

**Competing interests:** The authors declare no competing interests.

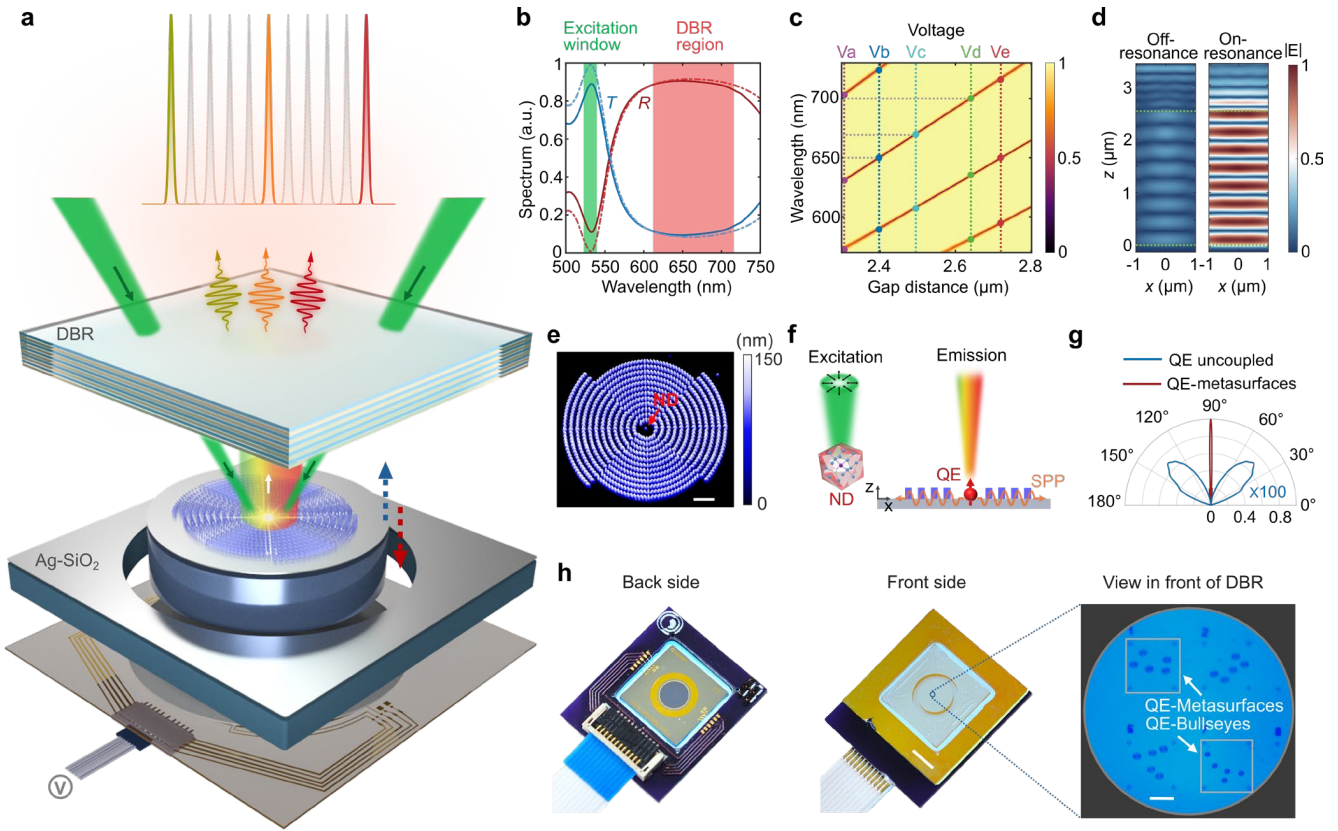

**Fig. 1 | MEMS-QEMS integrated platform for tunable photon sources. a**, Schematic illustration of tunable solid-state photon sources enabled by a chip-scale platform integrating QEMS into a MEMS-DBR tunable microcavity. **b**, The reflectance ($R$) and transmittance ($T$) of the DBR. The solid lines represent the measured transmittance (blue) and reflectance (red), while the dashed lines correspond to the calculated results. **c**, Mode characterization of the tunable microcavity formed by the DBR and MEMS-actuated mirror. The cavity resonance wavelengths, measured when applying different voltages, are marked as dots on the dashed lines and match well with the calculation. **d**, The electric field distributions for off-resonance ($\lambda = 650$ nm) and on-resonance ($\lambda = 670$ nm) regimes under the voltage Vc in (**c**). **e**, An AFM image of the QEMS. The scale bar, 2 μm. **f**, Schematic illustration of the excitation and emission processes of QEMS. A radially polarized 532 nm laser is used to excite NV-centres in NDs. **g**, Angular distributions of the emission for QE uncoupled and coupled to metasurfaces. **h**, Typical photos (back and front sides) of the integrated tunable solid-state photon sources. The scale bar on the front side of the device, 3 mm. The microscope image is taken in front of DBR. The scale bar, 50 μm.
11

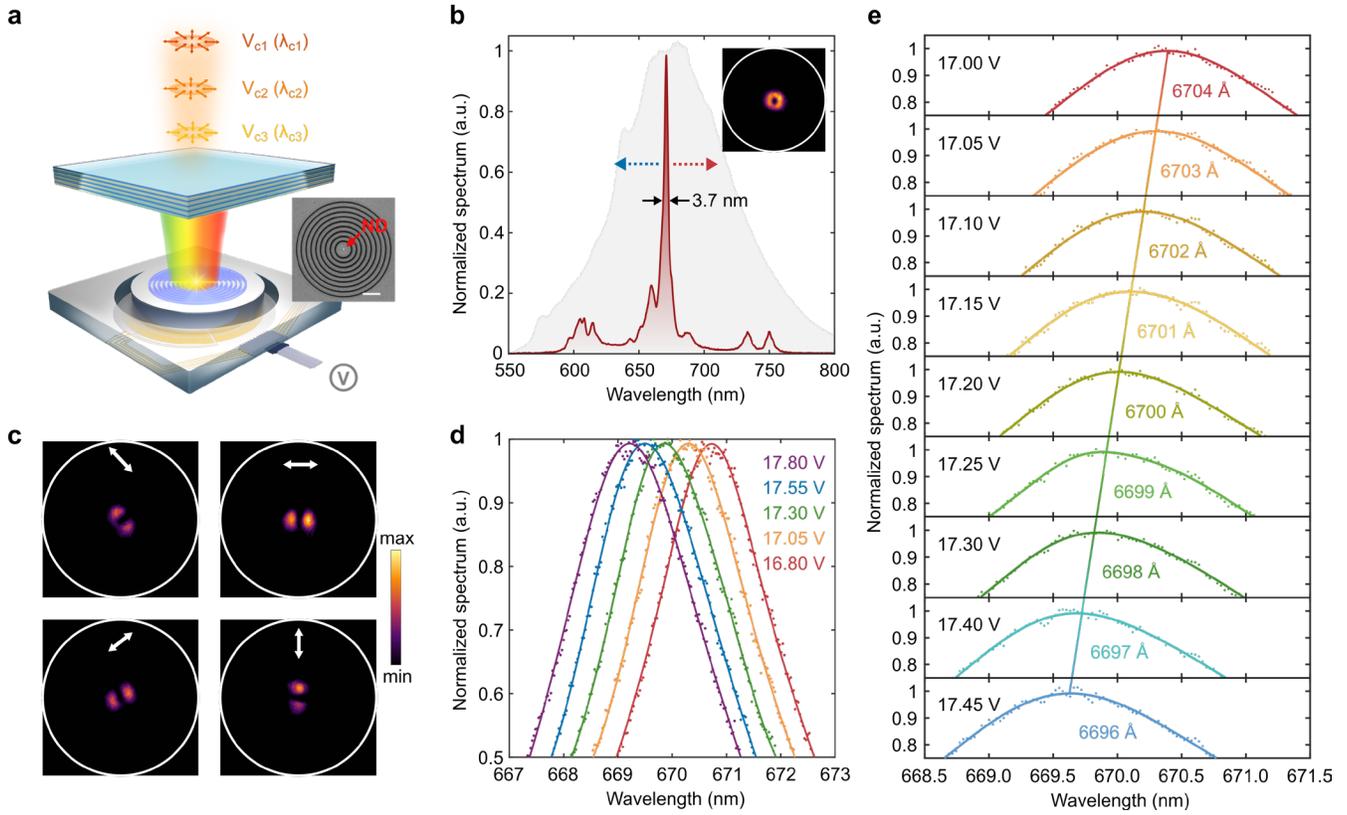

**Fig. 2 | Demonstration of narrow bandwidth and angstrom level tunability of photon emission. a**, Schematic configuration of QE-coupled Bullseye grating in the MEMS-DBR tunable cavity platform. The inset shows a scanning electron microscope (SEM) image, in which the ND position is marked. The scale bar, 2 μm. **b**, Narrow bandwidth photon emission of QE-coupled Bullseye grating in the MEMS-DBR cavity. The grey shaded area indicates the normalized spectrum of the QE-coupled Bullseye grating before integration into the MEMS-DBR microcavity. The inset is the far-field emission pattern, with the white circle indicates NA = 0.5. **c**, Far-field emission patterns measured while rotating a linear polarizer, demonstrating radially polarized photon emission. The white circles indicate NA = 0.5. **d,e**, Normalized photon emission spectra when varying voltages (**d**) in 250 mV steps and (**e**) with fine tuning. Dots are measured data, and the coloured lines are fitted curves.

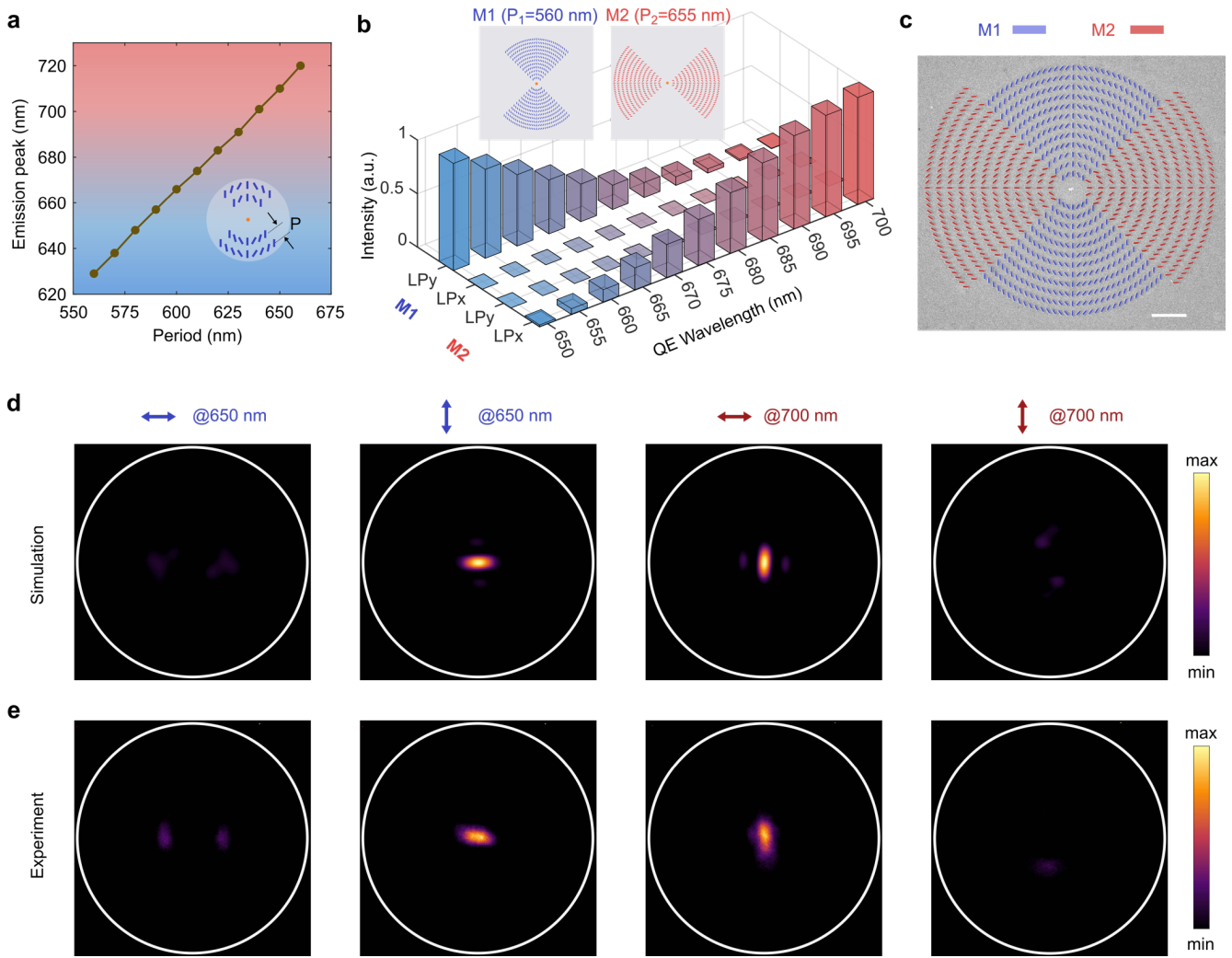

**Fig. 3 | Wavelength-polarization-multiplexed QEMS for photon emission of different polarizations encoded in different wavelengths. a**, The relationship between the metasurface period and the peak wavelength of outcoupled photon emission. The inset shows the first two periods of QEMS. The orange spot in the centre indicates the QE position. **b**, The performances of two metasurfaces (M1 and M2) with different periods ($P_1$ = 560 nm and $P_2$ = 655 nm), targeting for photon emission at $\lambda$ = 650 nm (with LPy state) and $\lambda$ = 700 nm photon emission (with LPx state), respectively. **c**, Colorized SEM image of the wavelength-polarization-multiplexed QEMS with the composite M1 and M2 parts. The scale bar, 2 μm. **d,e**, Far-field emission patterns at $\lambda$ = 650 nm and $\lambda$ = 700 nm, (**d**) simulation and (**e**) experiment results, the arrows at the top indicating different orientations of the linear polarization analyser. The white circles indicate NA = 0.5.

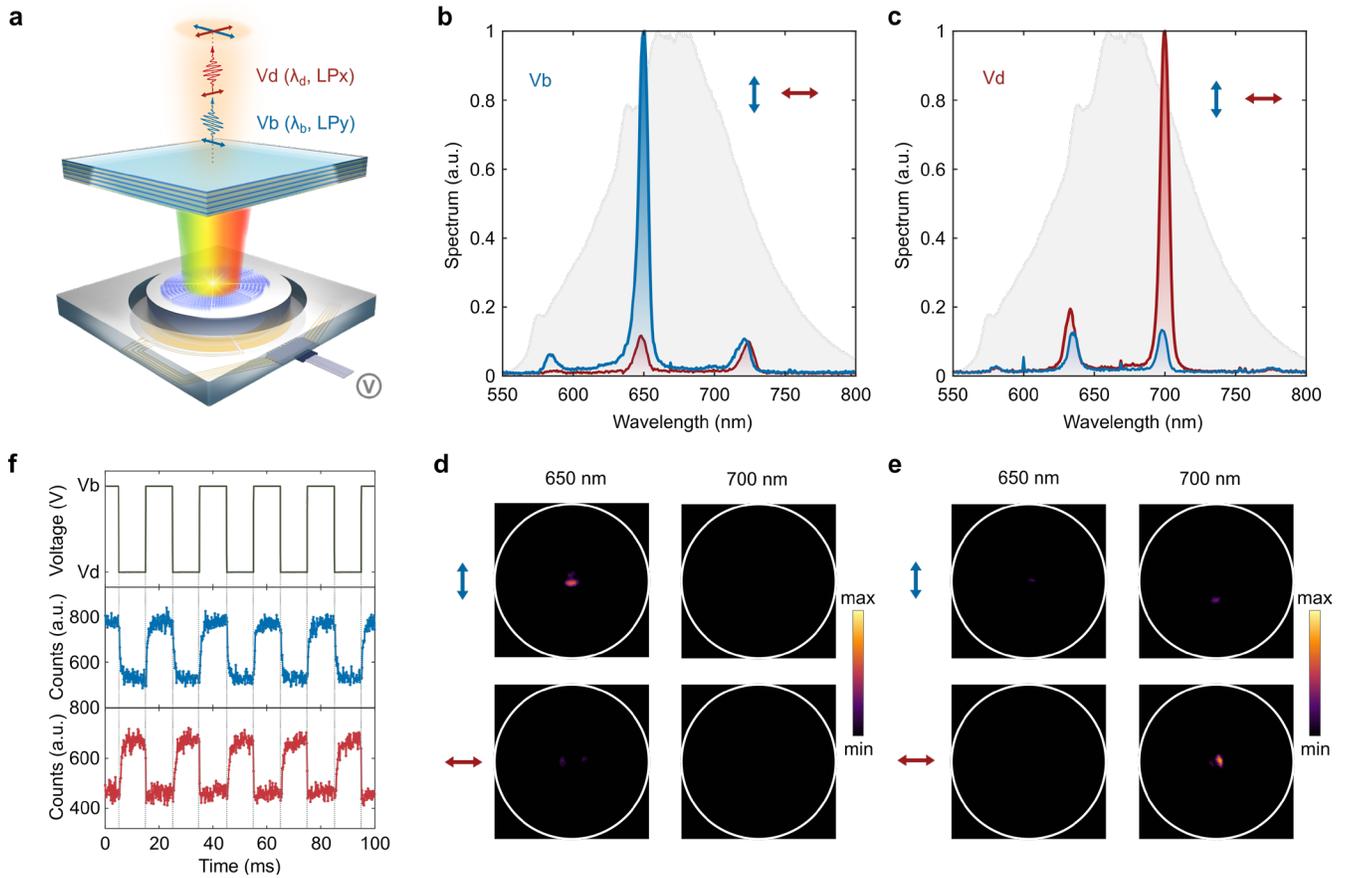

**Fig. 4 | Dynamic photon sources with wide range tunability and polarization switching. a**, Schematic configuration of the dynamic polarized solid-state photon sources, integrating multiplexed QEMS in the MEMS-DBR microcavity **b,c**, Measured photon emission spectra with LPx and LPy polarizations under different actuation voltages (**b**) Vb and (**c**) Vd. The grey shaded area indicates the normalized spectrum of QEMS before integration into the MEMS-DBR microcavity. The blue line and arrow correspond to LPy, and the red line and arrow correspond to LPx. **d,e**, Measured far-field emission patterns at $\lambda =$ 650 nm and $\lambda =$ 700 nm with different polarizations by applying voltages of (**d**) Vb and (**e**) Vd. The white circles denote NA = 0.5. **f**, Response time for switching between two operation states, measured by actuating the MEMS mirror with a periodic rectangle signal and detecting photons at different wavelengths.

**References**

1. Aharonovich, I., Englund, D. & Toth, M. Solid-state single-photon emitters. *Nat. Photonics* **10**, 631–641 (2016).

2. Li, Z. *et al.* Atomic optical antennas in solids. *Nat. Photonics* **18**, 1113–1120 (2024)

3. Wang, J., Sciarrino, F., Laing, A. & Thompson, M. G. Integrated photonic quantum technologies. *Nat. Photonics* **14**, 273–284 (2020).

4. Sun, S., Kim, H., Luo, Z., Solomon, G. S. & Waks, E. A single-photon switch and transistor enabled by a solid-state quantum memory. *Science* **361**, 57–60 (2018).

5. Utzat, H. *et al.* Coherent single-photon emission from colloidal lead halide perovskite quantum dots. *Science* **363**, 1068–1072 (2019).

6. Jelezko, F. & Wrachtrup, J. Single defect centres in diamond: A review. *Physica Status Solidi (a)* **203**, 3207–3225 (2006).

7. Ates, S. *et al.* Non-resonant dot–cavity coupling and its potential for resonant single-quantum-dot spectroscopy. *Nat. Photonics* **3**, 724–728 (2009).

8. Liu, S. *et al.* Super-resolved snapshot hyperspectral imaging of solid-state quantum emitters for high-throughput integrated quantum technologies. *Nat. Photonics* **18**, 967–974 (2024).

9. Ma, J. *et al.* Engineering quantum light sources with flat optics. *Adv. Mater.* **36**, 2313589 (2024).

10. Kan, Y. & Bozhevolnyi, S. I. Advances in metaphotonics empowered single photon emission. *Adv. Opt. Mater.* **11**, 2202759 (2023).

11. Koenderink, A. F. Single-photon nanoantennas. *ACS Photonics* **4**, 710–722 (2017).

12. Chen, B. *et al.* Bright solid-state sources for single photons with orbital angular momentum. *Nat. Nanotechnol.* **16**, 302–307 (2021).

13. Javadi, A. et al. Spin–photon interface and spin-controlled photon switching in a nanobeam waveguide. *Nat. Nanotechnol.* **13**, 398–403 (2018).

14. Akselrod, G. M. et al. Probing the mechanisms of large Purcell enhancement in plasmonic nanoantennas. *Nat. Photonics* **8**, 835–840 (2014).

15. Benz, F. et al. Single-molecule optomechanics in "picocavities." *Science* **354**, 726–729 (2016).

16. Bogdanov, S. I. *et al.* Ultrabright room-temperature sub-nanosecond emission from single nitrogen-vacancy centers coupled to nanopatch antennas. *Nano Lett.* **18**, 4837–4844 (2018).

17. Kan, Y. *et al.* Metasurface-enabled generation of circularly polarized single photons. *Adv. Mater.* **32**, 1907832 (2020).

18. Liu, X. *et al.* On-chip generation of single-photon circularly polarized single-mode vortex beams. *Sci. Adv.* **9**, eadh0725 (2023).

19. Komisar, D., Kumar, S., Kan, Y., Wu, C. & Bozhevolnyi, S. I. Generation of radially polarized single photons with plasmonic bullseye antennas. *ACS Photonics* **8**, 2190–2196 (2021).

20. Liu, X. *et al.* Ultracompact single-photon sources of linearly polarized vortex beams. *Adv. Mater.* **36**, 2304495 (2024).

21. Liu, X. *et al.* Off-normal polarized single-photon emission with anisotropic holography metasurfaces. *Nano Lett.* **24**, 13867–13873 (2024).

22. Ha, S. T. *et al.* Optoelectronic metadevices. *Science* **386**, eadm7442 (2024).

23. Forbes, A., De Oliveira, M. & Dennis, M. R. Structured light. *Nat. Photonics* **15**, 253–262 (2021).

24. Nape, I., Sephton, B., Ornelas, P., Moodley, C. & Forbes, A. Quantum structured light in high dimensions. *APL Photon.* **8**, 051101 (2023).

25. Nowak, A. K. *et al.* Deterministic and electrically tunable bright single-photon source. *Nat. Commun.* **5**, 3240 (2014).

26. Larocque, H. *et al.* Tunable quantum emitters on large-scale foundry silicon photonics. *Nat. Commun.* **15**, 5781 (2024).

27. Albrecht, R., Bommer, A., Deutsch, C., Reichel, J. & Becher, C. Coupling of a single nitrogen-vacancy center in diamond to a fiber-based microcavity. *Phys. Rev. Lett.* **110**, 243602 (2013).

28. Casabone, B. *et al.* Dynamic control of Purcell enhanced emission of erbium ions in nanoparticles. *Nat. Commun.* **12**, 3570 (2021).

29. Xia, K. *et al.* Tunable microcavities coupled to rare-earth quantum emitters. *Optica* **9**, 445–450 (2022).

30. Yang, J. *et al.* Tunable quantum dots in monolithic Fabry-Perot microcavities for high-performance single-photon sources. *Light Sci. Appl.* **13**, 33 (2024).

31. Tomm, N. *et al.* A bright and fast source of coherent single photons. *Nat. Nanotechnol.* **16**, 399–403 (2021).

32. Ding, X. *et al.* High-efficiency single-photon source above the loss-tolerant threshold for efficient linear optical quantum computing. *Nat. Photonics* **19**, 387–391 (2025).

33. Meng, C. *et al.* Dynamic piezoelectric MEMS-based optical metasurfaces. *Sci. Adv.* **7**, eabg5639 (2021).

34. Meng, C., Thrane, P. C., Ding, F. & Bozhevolnyi, S. I. Full-range birefringence control with piezoelectric MEMS-based metasurfaces. *Nat. Commun.* **13**, 2071 (2022).

35. Ding, F., Meng, C. & Bozhevolnyi, S. I. Electrically tunable optical metasurfaces. *Photonics Insights* **3**, R07–R07 (2024).

36. Andersen, S. K. H. *et al.* Hybrid Plasmonic Bullseye Antennas for Efficient Photon Collection. *ACS Photonics* **5**, 692–698 (2018).

37. Bakke, T. *et al.* A novel ultra-planar, long-stroke and low-voltage piezoelectric micromirror. *J. Micromech. Microeng.* **20**, 064010 (2010).

38. Kan, Y. *et al.* High-dimensional spin-orbital single-photon sources. *Sci. Adv.* **10**, eadq6298 (2024).

39. Tang, H. *et al.* On-chip multi-degree-of-freedom control of two-dimensional materials. *Nature* **632**, 1038–1044 (2024).

40. Dahl-Hansen, R., Gjessing, J., Mardilovich, P., Fragkiadakis, C. & Thorstensen, J. Reliable Pb (Zr, Ti) $O_3$-based thin film piezoelectric micromirrors for space-applications. *Appl. Phys. Lett.* **121**, 132901 (2022).

41. Kan, Y., Liu, X., Kumar, S. & Bozhevolnyi, S. I. Multichannel quantum emission with on-chip emitter-coupled holographic metasurfaces. *ACS Nano* **17**, 20308–20314 (2023).

42. Chen, O. *et al.* Compact high-quality CdSe–CdS core–shell nanocrystals with narrow emission linewidths and suppressed blinking. *Nat. Mater.* **12**, 445–451 (2013).

43. Iwasaki, T. *et al.* Tin-vacancy quantum emitters in diamond. *Phys. Rev. Lett.* **119**, 253601 (2017).

44. Jantzen, U. *et al.* Nanodiamonds carrying silicon-vacancy quantum emitters with almost lifetime-limited linewidths. *New J. Phys.* **18**, 073036 (2016).

45. Asbahi, M. *et al.* Large area directed self-assembly of sub-10 nm particles with single particle positioning resolution. *Nano Lett.* **15**, 6066–6070 (2015).

46. Pambudi, M. T. *et al.* Deterministic positioning of few aqueous colloidal quantum dots. *Nanoscale* **16**, 18339–18347 (2024).

47. Meng, C., Thrane, P. C., Wang, C., Ding, F. & Bozhevolnyi, S. I. MEMS-tunable topological bilayer metasurfaces for reconfigurable dual-state phase control. *Optica* **11**, 1556–1566 (2024).

48. Bartholomew, J. G., De Oliveira Lima, K., Ferrier, A. & Goldner, P. Optical line width broadening mechanisms at the 10 kHz level in $Eu^{3+}$: $Y_2 O_3$ nanoparticles. *Nano Lett.* **17**, 778–787 (2017).

49. Erhard, M., Krenn, M. & Zeilinger, A. Advances in high-dimensional quantum entanglement. *Nat. Rev. Phys.* **2**, 365–381 (2020).

50. Lukin, D. M. *et al.* Spectrally reconfigurable quantum emitters enabled by optimized fast modulation. *Npj Quantum Inf.* **6**, 80 (2020).

51. Pirandola, S. *et al.* Advances in quantum cryptography. *Adv. Opt. Photonics* **12**, 1012–1236 (2020).

## Methods

**Numerical simulations**

Three-dimensional (3D) numerical simulations of quantum emitter (QE) embedded metasurfaces (QEMS) were conducted using the finite-difference time-domain (FDTD) method. The QE was modelled as a *z*-direction electric dipole radiating at various wavelengths within the main emission region of nitrogen vacancy centres in nanodiamonds (NVs-ND). The dipole was placed 50 nm above a $SiO_2$ spacer (20 nm) covered silver film (150 nm) and in the centre of QEMS constructed by the dielectric material (with thickness of 150 nm, refractive index 1.41). A two-dimensional monitor atop the configuration was employed to obtain far-filed electric fields by near- to far-field transformation method.

Transfer matrix method (TMM) was used to check the performance of the microelectromechanical system (MEMS) positioned microcavity and estimate the air gap distances between the MEMS mirror and the distributed Bragg reflector (DBR). Experimentally measured refractive indices of $SiO_2$ and $TiO_2$, as shown in Supplementary Fig. 1, were utilized for the TMM calculations to ensure accuracy.

**Device fabrication**

The device fabrication process consists of three main parts: fabrication of piezoelectric MEMS chips, fabrication of QEMS, and assembly with MEMS-DBR.

***Fabrication of piezoelectric MEMS chips.*** The MEMS chips were fabricated by depositing a 2 μm thick layer of the piezoelectric material PZT between two electrode layers (bottom Pt and top Au) onto an SOI wafer with an 8 μm thick device layer. The device layer side was then bonded to a support wafer using WaferBOND HT-10.1, this was done to protect the MEMS devices during dicing and later during spin coating processes while fabricating the QEMS. An annular trench was subsequently etched using deep reactive ion etching through the substrate and oxide layers of the SOI wafer, leaving a Si membrane with an electrode-PZT-electrode stack that can mechanically move the remaining Si central mass when applying actuation voltages across the piezoelectric PZT.

***Fabrication of QEMS.*** The fabrication of QEMS followed a well-established sequence of technological steps (Extended Data Fig. 1 top row). Firstly, a 150 nm-thick silver film was deposited by thermal evaporation onto MEMS chips from a 150 mm SOI wafer that is bonded to a support wafer, followed by a 20 nm-thick $SiO_2$ layer being deposited using magnetron sputtering. Secondly, a group of alignment markers were fabricated on the $SiO_2$-covered Ag film through a combined process, including electron beam lithography (EBL) with a JEOL-6490 system (30 kV accelerating voltage), gold film deposition, and lift-off. Thirdly, a solution of NVs-NDs was spin-coated onto the surface with a proper diluted concentration and spin speed. Then, the positions of NVs-NDs related to the prefabricated alignment markers were determined by the dark-field microscope image with the developed precise alignment procedure. A negative photoresist (hydrogen silsesquioxane, HSQ) was spin-coated on the surface at 3000 rpm for 45s and baked on a hotplate at 160 °C for 2 minutes, forming a ~150 nm-thick layer. The thickness was verified using an atomic force microscope (AFM, NT-MDT NTEGRA). Subsequently, a second round EBL was performed to pattern the designed QEMS around the NVs-NDs. The on-chip QEMS were developed using tetramethylammonium hydroxide solution for 4 min, followed by rinsing in water for 60 s. The scanning electron microscopy (SEM) images were obtained with a 30-kV JEOL-6490 electron microscope.

***Assembly with MEMS-DBR.*** After QEMS fabrication, the support chip was removed by dissolving the waferBOND with WaferBOND Remover. The DBR was fabricated using reactive DC sputtering, with $SiO_2$ layers deposited from a Si target and $TiO_2$ layers deposited from a Ti target in oxygen atmosphere. The DBR was then mounted onto the structured MEMS mirror using Loctite 401. Finally, the device was assembled with a printed circuit board (PCB) and wire-bonded for electrical connections (Extended Data Fig. 1 bottom row).

**Optical characterization**

***Characterization of the MEMS-DBR microcavities.*** The MEMS-DBR microcavities were characterized by measuring the normalized reflection spectra[33,34], as shown in Supplementary Fig. 3. A halogen lamp (KL1500 LCD, Schott, wavelength range: ~400-1000 nm) was used as the broadband light source. Two Irises (ID1: Iris 1, SM1D12SZ, Thorlabs and ID2: Iris 2, ID25, Thorlabs) were positioned at the first image and Fourier plane to filter the area of interest in the measurement. The Flip lens (LBF254-100-A, Thorlabs) was used to switch between direct imaging and Fourier imaging. The VIS-NIR spectrometer (QE pro, Ocean Optics) was used to take spectra. The reflection from MEMS-DBR tunable microcavity was normalized to the reflection from the MEMS mirror. The actuation voltages applied to the MEMS can be controlled in real time using a computer-connected controller.

***Characterization of the QEMS and MEMS-QEMS devices.*** The QEMS and MEMS-QEMS devices were characterized using the experimental setup illustrated in Extended Data Fig. 2. A 532 nm laser beam (continuous wave or pulsed) was used to excite NVs-NDs after passing through a single-mode polarization-maintaining fiber and a radial polarization converter (ARCoptix RPC) to covert the linear polarization into radial polarization. The excitation and collection of photon emission were by the same objective with NA = 0.9 (The Olympus MPLFLN ×100). For measuring MEMS-QEMS, the objective NA = 0.7 (The Olympus MPLFLN ×50) was used for easily finding the QEMS beneath the DBR. The MEMS-QEMS device was mounted on a piezo stage for fluorescence imaging. The photon counts were recorded by the avalanche photo diodes (APDs). Fluorescence spectra were measured using the spectrometers (Andor Ultra 888 USB3 –BV; Ocean Optics QE pro) operating within the 550-800 nm range. Fluorescence decay-rate measurements were conducted using time-correlated single-photon counting (TCSPC) with a PicoHarp 300 TCSPC module correlating signals from both pulsed laser (Pico Quant LDH-PFA-530L) and PicoQuant τ-SPAD APD. The emission patterns were measured with rotating or without a linear polarizer by capturing Fourier plane images using a Hamamatsu Orca LT + CCD camera.

**Data availability**

All data that support the findings of the study are provided in the main text, its Extended Data, and Supplementary Information. Data are also available from the corresponding authors upon reasonable request.